\documentclass[a4paper]{article}
\usepackage{ISCSLP2026}
\usepackage{ifthen}
\usepackage{url}
\newboolean{blind}
\setboolean{blind}{false} 
\title{SyncVoice: Simple and Effective Automatic Video Dubbing with Vision-Augmented TTS}
\name{
	\ifthenelse{\boolean{blind}}{Anonymous to ISCSLP}
	{Kaidi Wang$^1$,Yi He$^3$,Wenhao Guan$^2$,Weijie Wu$^1$,Peijie Chen$^1$,Hongwu Ding$^3$,Xiong Zhang$^3$,Di Wu$^3$$^,$$^4$,Meng Meng$^3$,Jian Luan$^3$,Lin Li$^2$$^\dag$,Qingyang Hong$^1$$^\dag$}
}
\address{
  \ifthenelse{\boolean{blind}}{Anonymous to ISCSLP}
  {
  	$^1$School of Informatics, Xiamen University, China\\
  $^2$School of Electronic Science and Engineering, Xiamen University, China\\
  $^3$MiLM Plus, Xiaomi Inc., China\\
  $^4$WeNet Open Source Community
  }
  \thanks{\dag\ Corresponding author.}
}

\email{
	\ifthenelse{\boolean{blind}}{Anonymous to ISCSLP}
	{kaidi@stu.xmu.edu.cn}
}

\begin{document}

\maketitle
\begin{abstract}
Automatic video dubbing aims to generate high-fidelity speech that is temporally aligned with visual content. However, existing methods still suffer from limited speech naturalness, insufficient audio-visual synchronization, and poor scalability beyond monolingual settings. To address these challenges, we propose SyncVoice, a simple and effective dubbing framework that lightly integrates a Text-Visual Fusion Module into a pretrained text-to-speech (TTS) system. This module aligns visual features with linguistic representations, enabling temporally synchronized speech synthesis without complex architectural redesign. Experiments on the LRS3 dataset show that SyncVoice achieves state-of-the-art performance in zero-shot dubbing. Further training on a large-scale bilingual audio-visual dataset improves vocal fidelity while preserving synchronization, yielding a single unified model for both Chinese and English dubbing.

\end{abstract}

\section{Introduction}

In recent years, the rapid growth of short-video platforms and global content sharing has significantly increased the demand for automatic video dubbing \cite{hu2021neural,9746421}. This dubbing task aims to generate natural and expressive speech from text scripts while preserving a target speaker's timbre. At the same time, the generated speech must be strictly synchronized with visual cues such as lip movements in the input video. Unlike conventional text-to-speech (TTS) systems \cite{ren2019fastspeech}, automatic video dubbing additionally requires modeling visual information, making it a challenging problem at the intersection of speech synthesis and multimodal learning.

Early dubbing approaches \cite{hu2021neural} extract lip movement features to guide speech generation and utilize facial representations for speaker modeling in multi-speaker scenarios. HPMDubbing \cite{cong2023learning} further proposes a hierarchical prosody modeling framework that incorporates visual cues from lips, face, and scene context, while extracting speaker representations from reference audio to improve voice consistency. However, these models \cite{hu2021neural,cong2023learning,cong-etal-2024-styledubber,Cong2024EmoDubberTH} are typically trained on relatively small audio-visual datasets, which limits their generalization ability. To address this limitation, several recent studies \cite{Zhang2024FromST, Zhang2025ProsodyEnhancedAP, Choi2025AlignDiTMA, sung2025voicecraft} leverage large-scale speech corpora or pretrained TTS models \cite{Peng2024VoiceCraftZS,Eskimez2024E2TE,Chen2024F5TTSAF}, followed by fine-tuning on audio-visual data. While these approaches significantly improve speech naturalness, challenges remain in achieving precise audio-visual synchronization, and most existing systems are restricted to single-language settings.

In this work, we propose SyncVoice, a novel automatic video dubbing framework built upon a pretrained TTS model. Instead of training from scratch—which is often constrained by limited and noisy audio-visual data \cite{Zhang2024FromST}—our method leverages ZipVoice \cite{zhu2025zipvoice}, a flow-matching-based TTS model with strong zero-shot capabilities. We further introduce a Text-Visual Fusion Module that aligns textual and visual representations over time, enabling the generation of speech that is temporally consistent with the speaker's lip movements. This design allows the model to maintain the speech quality of the pretrained backbone while improving audio-visual synchronization.
Our main contributions are summarized as follows:

\begin{itemize}
    \item We propose SyncVoice, a simple non-autoregressive automatic video dubbing framework based on flow matching, which augments a pretrained TTS backbone with lightweight multimodal conditioning for synchronized, high-fidelity speech synthesis.
    \item We introduce an early multimodal fusion strategy that injects visual cues at the linguistic level of a pretrained TTS model, achieving effective audio-visual synchronization without complex fusion modules or auxiliary alignment objectives.
    \item Extensive experiments on LRS3 show that SyncVoice achieves state-of-the-art zero-shot dubbing performance, and scaling to a large-scale Chinese--English bilingual corpus further demonstrates strong generalization.\footnote{Demos are available at \url{https://wkd88.github.io/syncvoice/}.}
\end{itemize}


\section{Methodology}
\subsection{Model Architecture}
\begin{figure}[htb]
\centering
\includegraphics[width=0.9\linewidth]{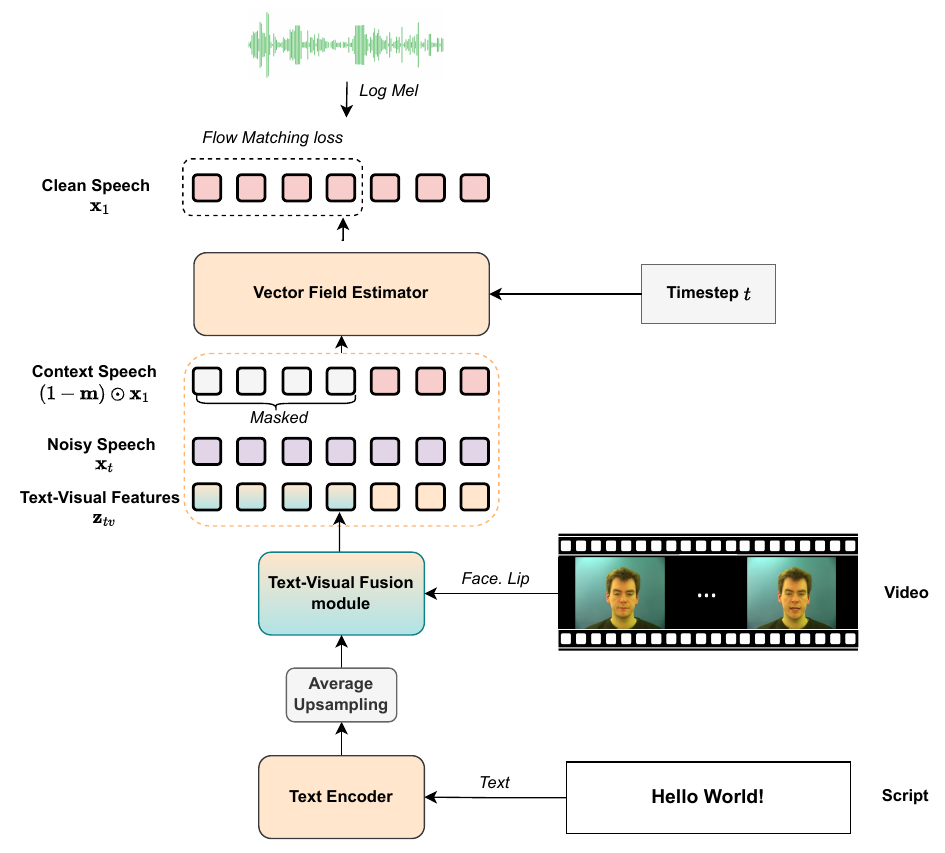}
\caption{The main architecture of the proposed method.}
\label{fig:fig1}
\end{figure}

Fig~\ref{fig:fig1} presents the overall architecture of SyncVoice. 
It consists of two components: 
(1) a non-autoregressive TTS backbone, 
(2) a Text-Visual Fusion module, and 
We describe each component below.

\subsubsection{TTS Backbone}

We adopt ZipVoice~\cite{zhu2025zipvoice} as the backbone of our system. 
ZipVoice is a non-autoregressive TTS model based on the Flow Matching paradigm, 
consisting of a text encoder that maps token sequences into contextual linguistic 
representations and a vector field estimator that models continuous latent trajectories 
for high-quality mel-spectrogram generation.

A key component that distinguishes ZipVoice from many recent non-autoregressive TTS models \cite{Eskimez2024E2TE,Chen2024F5TTSAF} 
is the Average Upsampling mechanism. Instead of padding text tokens with 
additional filler tokens to match the acoustic length, the model expands 
linguistic representations to the acoustic frame rate by uniformly repeating 
each token according to the ratio between the target acoustic length and the 
text sequence length. This simple yet effective strategy establishes coarse 
temporal alignment between linguistic and acoustic representations at an early 
stage, which accelerates the learning of text–speech alignment and improves 
training convergence. Furthermore, the resulting frame-level linguistic 
representation provides a natural interface for incorporating additional 
temporal signals, which is particularly advantageous when integrating 
multimodal cues such as visual information.

\begin{figure}[htb]
\centering
\includegraphics[width=0.6\linewidth]{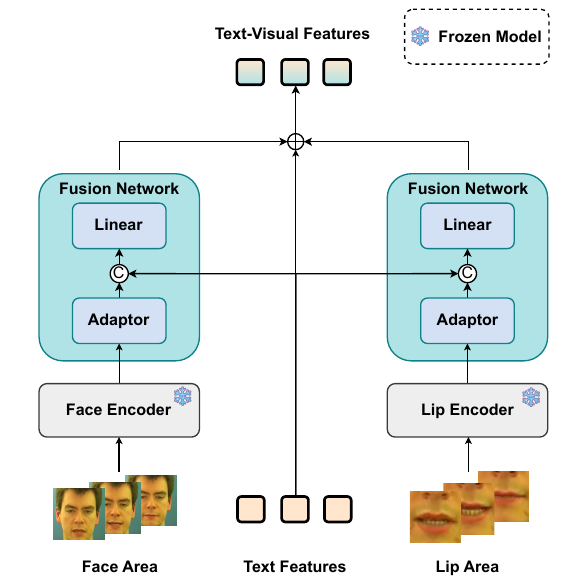}
\caption{Detail of Text-Visual Fusion module.}
\label{fig:fig2}
\end{figure}

\subsubsection{Text-Visual Fusion Module}

Building on the aligned representation space provided by Average Upsampling, 
we introduce a Text-Visual Fusion Module to incorporate visual cues for improved 
audio–visual synchronization, as illustrated in Fig.~\ref{fig:fig2}. 
Following prior dubbing works~\cite{cong-etal-2024-styledubber,sung2025voicecraft}, 
we employ both facial and lip features as visual conditions. 
Face regions are first detected and cropped from each video frame using RetinaFace 
\cite{9157330}, and visual representations are extracted using pretrained visual 
encoders~\cite{toisoul2021estimation,martinez2020lipreading}.

To ensure temporal consistency across modalities, the visual feature sequence is 
interpolated to match the acoustic frame resolution and the upsampled text sequence. 
The aligned visual features are then fused with the upsampled text representations 
through a lightweight fusion network based on residual adaptation. Specifically, 
visual representations are first transformed by a bottleneck adapter and then 
concatenated with textual features, followed by a linear projection to the model 
hidden dimension. The final resulting representation is referred to as the 
\textit{Text-Visual Features}, which serve as the multimodal condition for the 
subsequent generation module.

By injecting facial and lip cues into the linguistic representation space, 
the proposed module enables the model to capture visual speech dynamics and 
achieve more stable audio–visual synchronization. Compared with recent 
flow matching-based dubbing method that relies on complex multimodal fusion strategies 
and auxiliary alignment objectives~\cite{Choi2025AlignDiTMA}, as well as 
autoregressive approach that generates speech sequentially~\cite{sung2025voicecraft}, 
our method leverages the structural alignment induced by Average Upsampling 
to enforce temporal consistency. This design reduces the reliance on auxiliary 
alignment losses and mitigates cross-modal drift, leading to more stable 
multimodal dubbing.

\begin{table*}[htbp]
    \centering
    \caption{Results on the LRS3 dataset. Our method incorporates visual cues (facial and lip features) and uses multi-condition CFG with scaling factors \(s_t=0.8, s_f=0.0, s_l=0.5\). The best results are shown in bold and the second-best are underlined.}
    \label{tab:2}
    \resizebox{0.75\textwidth}{!}{
    \begin{tabular}{l c c c c c c c c}
        \toprule
        Methods & LSE-C $\uparrow$ & LSE-D $\downarrow$ & WER $\downarrow$ & SIM-o $\uparrow$ & UTMOS $\uparrow$ & MCD $\downarrow$ & MOS-N $\uparrow$ & MOS-S $\uparrow$ \\
        \midrule
        Ground Truth & 7.33 & 7.67 & 1.94 & 0.783 & 3.87 & - & 4.42 $\pm$ 0.12 & 4.21 $\pm$ 0.09 \\
        AlignDiT \cite{Choi2025AlignDiTMA} & \underline{7.79} & \underline{7.19} & \textbf{2.26} & \underline{0.681} & 3.69 & \underline{6.36} & \underline{4.12 $\pm$ 0.14} & \underline{4.01 $\pm$ 0.13} \\
        VoiceCraft-Dub \cite{sung2025voicecraft} & 6.64 & 8.26 & 11.49 & 0.535 & \underline{3.75} & 6.81 & 4.06 $\pm$ 0.10 & 3.53 $\pm$ 0.12  \\
        \midrule
        \textbf{SyncVoice} & \textbf{8.36} & \textbf{6.83} & \underline{2.40} & \textbf{0.685} & \textbf{3.98} & \textbf{6.35} & \textbf{4.23 $\pm$ 0.11} & \textbf{4.03 $\pm$ 0.12} \\
        \bottomrule
    \end{tabular}
    }
\end{table*}

\subsection{Training Method}

We train the model using a masked Flow Matching \cite{Lipman2022FlowMF} formulation. 
Let $\mathbf{x}_1 \in \mathbb{R}^{D \times T}$ denote the ground-truth speech spectrogram. 
During training, a binary mask $\mathbf{m} \in \{0,1\}^{D \times T}$ is randomly sampled, 
where $m_{d,t}=1$ indicates a masked element to be generated at frequency bin $d$ 
and time frame $t$, while $m_{d,t}=0$ denotes an observed context element. 
A diffusion time step $t \sim \mathcal{U}(0,1)$ is sampled and embedded before being fed 
into the Vector Field Estimator parameterized by $\theta$. 
The estimator takes as input the fused text–visual features $\mathbf{z}_{\text{tv}}$, 
the observed speech context $(1-\mathbf{m}) \odot \mathbf{x}_1$, and the interpolated noisy speech

\begin{equation}
\mathbf{x}_t = (1-t)\mathbf{x}_0 + t\mathbf{x}_1,
\end{equation}
where $\mathbf{x}_0 \sim \mathcal{N}(0, I)$ is Gaussian noise and $\odot$ denotes element-wise multiplication.

During training, visual features are provided only for masked regions, while non-masked regions are replaced with masked visual features. 
This design enables inference using only reference audio without requiring reference video, improving the practicality of the system. Finally, the model is optimized using the following masked Flow Matching objective:

\begin{equation}
\begin{aligned}
\mathcal{L}_{\text{FM-Dub}}
= \mathbb{E}_{t, \mathbf{x}_1, \mathbf{x}_0}
\Big[
\big|\big|
\mathbf{m} \odot
\big(
\mathbf{v}_t(\mathbf{x}_t, \mathbf{z}_{\text{tv}},
(1 - \mathbf{m}) \odot \mathbf{x}_1; \theta)
\\
- (\mathbf{x}_1 - \mathbf{x}_0)
\big)
\big|\big|_2^2
\Big],
\end{aligned}
\label{eq:fm_loss}
\end{equation}

\subsection{Multi-Condition Classifier-Free Guidance}

We adopt a multi-condition classifier-free guidance (CFG) strategy \cite{ho2021classifierfree} 
to enable controllable speech generation from multiple modalities. 
The model is conditioned on facial features ($f$), lip features ($l$), and text ($t$). 
During training, each condition is randomly dropped with predefined probabilities, 
and when all conditions are removed the model receives an unconditional input, 
denoted as $\varnothing$.

At inference time, predictions under different conditioning configurations 
are combined to control the influence of each modality. 
The final guided velocity field is computed as

\begin{equation}
\label{eq:cfg}
\mathbf{v}_{\mathrm{cfg}} = 
\mathbf{v}_{f,l,t} 
+ s_f \big( \mathbf{v}_{f,t} - \mathbf{v}_{t} \big)
+ s_l \big( \mathbf{v}_{l,t} - \mathbf{v}_{t} \big)
+ s_t \big( \mathbf{v}_{t} - \mathbf{v}_{\varnothing} \big),
\end{equation}

where $\mathbf{v}_{f,l,t}$ denotes the prediction under full conditioning, 
$\mathbf{v}_{f,t}$ and $\mathbf{v}_{l,t}$ denote predictions with the lip and facial modalities removed, respectively, and $\mathbf{v}_{\varnothing}$ denotes the unconditional prediction. 
The guidance scales $s_f$, $s_l$, and $s_t$ control the contribution of facial, lip, and textual conditions, respectively.


\section{Experimental Setup}

\subsection{Dataset}

We train two variants of our model under different data scales.

\textbf{(1) \textit{LRS3} (In-the-wild Zero-shot Dubbing):} 
The first variant is trained on the public LRS3-TED dataset~\cite{Afouras2018LRS3TEDAL}, 
which contains 439 hours of English speech from thousands of TED/TEDx speakers. 
Following the EMILIA protocol~\cite{he2024emilia}, 
we construct an EN-EN zero-shot test set 
by re-segmenting 1,054 video pairs from the original HDTF videos~\cite{zhang2021flow}. 
For each test sample, the reference speech is randomly selected 
from a different utterance of the same speaker.

\textbf{(2) \textit{Internal Bilingual Dataset} (Large-scale Bilingual Zero-shot Dubbing):} 
To investigate performance in a larger-scale bilingual scenario, 
we train a second model on an internal multi-speaker dataset 
comprising approximately 600 hours of Chinese 
and 1,190 hours of English video data, 
covering a broad speaker population and yielding substantially greater training diversity and scale.
In addition to the EN-EN test set, 
we construct a ZH-ZH zero-shot dubbing test set 
by randomly selecting 826 video pairs from CN-CVS \cite{10095796}. 
For each sample, the reference speech is randomly drawn 
from another utterance of the same speaker.

\subsection{Implementation Details}

Our model builds upon the ZipVoice architecture, consisting of a Zipformer-based text encoder and a vector field estimator. The text encoder contains 4 Zipformer layers with an embedding dimension of 192 and a feedforward dimension of 512. The vector field estimator adopts a 5-stack multi-scale Zipformer structure operating at downsampling rates of $[1\times, 2\times, 4\times, 2\times, 1\times]$ with $[2, 2, 4, 4, 4]$ layers, respectively, where each layer has an embedding dimension of 512 and a feedforward dimension of 1536. In the text-visual fusion module, each Adapter comprises a LayerNorm layer followed by a bottleneck structure (down-projection to $d/4$, GELU activation, and up-projection back to $d$), with a residual connection, an additional LayerNorm, and a linear layer. 
The final linear layer projects the concatenated features to a 100-dimensional representation.

The model is initialized with pretrained ZipVoice weights and optimized using Scaled Adam with a learning rate of $5 \times 10^{-4}$. 
All video clips are resampled to 25 FPS (frames per second). 
Audio and text processing follow the procedure in~\cite{zhu2025zipvoice}. We train the model for 55k steps on the LRS3 dataset using 4 NVIDIA H800 GPUs, and for 190k steps on the internal bilingual dataset using 2 NVIDIA RTX 5090 GPUs, with each GPU processing approximately 250 seconds of speech audio per step. 
During training, text features are randomly dropped with a probability of 0.2, while facial and lip features are each dropped with a probability of 0.6. 
During inference, we adopt an Euler ODE solver with 16 sampling steps. A pretrained vocoder \cite{Siuzdak2023VocosCT} is used to convert the generated speech features into waveforms.


\subsection{Evaluation Metrics}

\textbf{Subjective Evaluation.}
We randomly sample 30 utterances from the test set for human evaluation to assess the perceptual quality of the synthesized speech. 
Each sample is independently rated by 20 participants, who are instructed to score based on their subjective perception using a five-point Likert scale. 
The rating scale is designed such that 1 corresponds to "very poor" and 5 to "very good," with intermediate values representing varying levels of quality. 
Two criteria are evaluated during this subjective assessment: 
(1) Naturalness (MOS-N), which measures how natural and lifelike the synthesized speech sounds, 
and (2) Speaker Similarity (MOS-S), which quantifies how closely the synthesized voice matches the original speaker in terms of voice characteristics.

\textbf{Objective Evaluation.}
Objective performance is measured along five dimensions.
\textit{Speaker Similarity (SIM-o):}
We compute cosine similarity between speaker embeddings extracted from the reference speech and the synthesized audio. 
Embeddings are obtained using a WavLM-based ECAPA-TDNN \cite{Desplanques2020ECAPATDNNEC}.
\textit{Perceptual Quality (UTMOS):}
UTMOS \cite{Saeki2022UTMOSUS} is adopted as an automatic proxy for perceived speech naturalness.
\textit{Intelligibility (WER):}
Word Error Rate is calculated between the ground-truth text and ASR transcriptions of the speech. 
Whisper-large-v3 \cite{pmlr-v202-radford23a} is used for English evaluation, and Paraformer-zh \cite{Gao2022ParaformerFA} for Chinese.
\textit{Spectral Distortion (MCD):}
Mel-Cepstral Distortion is reported to quantify frame-level spectral deviation between generated and ground-truth mel-spectrograms.
\textit{Audio-Visual Synchronization (LSE-C / LSE-D):}
Following~\cite{li2024latentsync}, we employ SyncNet~\cite{prajwal2020lip} 
to compute Lip Sync Error Confidence (LSE-C) and Lip Sync Error Distance (LSE-D).

\section{Results}
\label{sec:results}
\subsection{Comparative Study on Public Baselines}

In this section, we evaluate the performance of our monolingual model by comparing it to state-of-the-art open-source baselines.

We compare SyncVoice with VoiceCraft-Dub~\cite{sung2025voicecraft} and AlignDiT~\cite{Choi2025AlignDiTMA}, both pretrained and evaluated on LRS3. As shown in Table~\ref{tab:2}, SyncVoice achieves the best results on most metrics. AlignDiT reports a marginally lower WER, largely due to an explicit CTC loss that enforces transcription accuracy; however, this auxiliary constraint comes at the cost of degraded performance in other dimensions, such as speaker similarity. By contrast, SyncVoice surpasses all baselines in SIM-o, UTMOS, MOS-N, MOS-S, LSE-C/LSE-D, and MCD. These results suggest that SyncVoice achieves more balanced dubbing quality in zero-shot settings with unseen speakers, improving synchronization and perceptual fidelity without sacrificing overall performance for WER alone.


\begin{table}[htbp]
    \centering
    \caption{Results on Bilingual Dataset (EN-EN test set). We set the CFG scaling factors as \(s_t = 1.0, s_f =0.0,s_l=0.5\).}
    \label{tab:2-2}
    \setlength{\tabcolsep}{4pt}
    \footnotesize
    \resizebox{0.45\textwidth}{!}{
    \begin{tabular}{l c c c c c c}
        \toprule
        Methods & LSE-C $\uparrow$ & LSE-D $\downarrow$ & WER $\downarrow$ & SIM-o $\uparrow$ & UTMOS $\uparrow$  & MCD $\downarrow$ \\
        \midrule
        Ground Truth & 7.33 & 7.67 & 1.94 & 0.783 & 3.87 & - 
        \\
        Zero-shot TTS & 2.86 & 12.06 & 2.39 & 0.733 & 4.08 & 6.42 \\
        \midrule
        \textbf{SyncVoice}$_{\mathrm{tts}}$ & 2.84 & 12.08 & \textbf{1.94} & 0.737 & 4.07 & 6.21 \\
        \textbf{SyncVoice}$_{\mathrm{face}}$ & 7.32 & 7.80 & \underline{1.98} & 0.735 & 4.07 & 5.84 \\
        \textbf{SyncVoice}$_{\mathrm{lip}}$ & \textbf{7.86} & \underline{7.30} & 2.08 & \textbf{0.738} & \textbf{4.08} & 5.75 \\
        \textbf{SyncVoice} & \textbf{7.86} & \textbf{7.29} & 2.03 & \underline{0.737} & \underline{4.07} & \textbf{5.73}\\ 
        \,\,w/ Random init. & 7.75 & 7.39 & 3.52 & 0.584 & 4.03 & 6.56 \\
        \bottomrule
    \end{tabular}
    }
\end{table}

\subsection{In-depth Exploration with Bilingual Data}

We further evaluate SyncVoice on a bilingual dataset to investigate its scalability across languages and multimodal inference settings. Results are reported on both English and Chinese test sets. We adopt the pretrained TTS model (Zero-shot TTS)~\cite{zhu2025zipvoice} as the baseline and adjust its output duration to match the video length under the default inference configuration. At inference time, SyncVoice$_{\text{tts}}$ performs text-only generation, while SyncVoice$_{\text{face}}$ and SyncVoice$_{\text{lip}}$ additionally use facial and lip features, respectively. SyncVoice (without a subscript) uses all available conditions. We also evaluate SyncVoice w/ Random Init., which is trained without pretrained initialization.

On the EN-EN test set (Table~\ref{tab:2-2}), SyncVoice$_{\text{tts}}$ largely preserves the synthesis capability of the pretrained TTS backbone while improving several metrics. Facial conditioning improves LSE-C, LSE-D, and MCD, although it slightly increases WER. Lip features provide stronger synchronization cues, yielding further improvements in synchronization and even surpassing Ground Truth on the synchronization-related metrics. They also marginally improve SIM-o and UTMOS while reducing MCD. Combining facial and lip features achieves the best LSE-D and MCD, suggesting that the two visual representations provide complementary information. Overall, visual conditioning improves synchronization while largely preserving speech quality for English dubbing. In contrast, SyncVoice w/ Random Init. consistently underperforms its pretrained-initialized counterpart, demonstrating the importance of pretrained TTS initialization.

On the ZH-ZH test set (Table~\ref{tab:2-3}), SyncVoice$_{\text{tts}}$ remains competitive with the pretrained baseline, achieving better SIM-o and MCD but slightly worse WER and UTMOS. Facial and lip conditioning progressively improves synchronization, while combining both features additionally improves WER. However, visual conditioning slightly reduces performance on some speech-quality metrics, and all visual configurations remain below Ground Truth in synchronization performance. This may be because the visual encoders and synchronization-related components are primarily trained on non-Chinese data and are less effective at capturing Chinese-specific articulation patterns. Incorporating more Chinese audiovisual data may therefore improve cross-lingual synchronization. SyncVoice w/ Random Init. again performs consistently worse across all metrics, confirming the importance of pretrained initialization.

Overall, scaling to large-scale bilingual data improves speech quality and demonstrates the strong generalization ability of SyncVoice for bilingual dubbing, while pretrained initialization is critical for achieving competitive performance across both languages.


\begin{table}[htbp]
    \centering
    \caption{Results on Bilingual Dataset (ZH-ZH test set). We set the CFG scaling factors as \(s_t = 0.8, s_f =0.0,s_l=0.5\).}
    \label{tab:2-3}
    \setlength{\tabcolsep}{4pt}
    \footnotesize
    \resizebox{0.45\textwidth}{!}{
    \begin{tabular}{l c c c c c c}
        \toprule
        Methods & LSE-C $\uparrow$ & LSE-D $\downarrow$ & WER $\downarrow$ & SIM-o $\uparrow$ & UTMOS $\uparrow$ & MCD $\downarrow$ \\
        \midrule
        Ground Truth & 5.34 & 8.44 & 0.89 & 0.709 & 2.68 & - \\
        Zero-shot TTS & 2.07 & 11.84 & 1.51 & 0.670 & 2.96 & 6.16 \\
        \midrule
        \textbf{SyncVoice}$_{\mathrm{tts}}$ & 1.97 & 11.92 & \textbf{1.72} & \textbf{0.682} & \textbf{2.85} & 6.13 \\
        \textbf{SyncVoice}$_{\mathrm{face}}$ & 2.56 & 11.33 & 1.99 & \underline{0.678} & \underline{2.80} & 6.12 \\
        \textbf{SyncVoice}$_{\mathrm{lip}}$ & \textbf{4.77} & \textbf{9.21} & 1.97 & 0.675 & 2.78 & \textbf{5.75} \\
        \textbf{SyncVoice} & \underline{4.75} & \underline{9.22} & \underline{1.91} & 0.674 & 2.76 & \underline{5.79} \\ 
        \,\,w/ Random init. & 4.57 & 9.28 & 11.22 & 0.590 & 2.57 & 5.85 \\
        \bottomrule
    \end{tabular}
    }
\end{table}

\begin{table}[htbp]
    \centering
    \caption{Ablation results for multi-condition CFG (EN-EN test set).}
    \setlength{\tabcolsep}{4pt}
    \label{tab:3}
    \footnotesize
    \resizebox{0.45\textwidth}{!}{
    \begin{tabular}{c c c | c c c c c c}
        \toprule
         \textbf{\(s_f\)} & \textbf{\(s_l\)} & \textbf{\(s_t\)} & LSE-C $\uparrow$ & LSE-D $\downarrow$ & WER $\downarrow$ & SIM-o $\uparrow$ & UTMOS $\uparrow$ & MCD $\downarrow$ \\
        \midrule
        1.0 & 1.0 & 1.0 & 8.02 & 7.15 & 2.24 & 0.719 & 4.04 & 6.13 \\
        0.5 & 0.5 & 1.0 & 7.92 & 7.23 & 2.20 & 0.733 & 4.07 & 5.83 \\
        0.5 & 0.5 & 1.5 & 7.79 & 7.35 & 1.90 & 0.730 & 4.05 & 6.02 \\
        0.5 & 0.5 & 0.5 & \textbf{8.03} & \textbf{7.14} & 2.38 & 0.722 & 4.06 & \textbf{5.72} \\
        0.5 & 0.0 & 1.0 & 7.69 & 7.45 & \textbf{1.84} & \textbf{0.738} & 4.07 & 5.75 \\
        0.0 & 0.5 & 1.0 & 7.86 & 7.29 & 2.03 & 0.737 & \textbf{4.07} & 5.73 \\
        0.0 & 0.0 & 1.0 & 6.42 & 8.49 & 4.38 & 0.734 & 3.47 & 6.12 \\
        \bottomrule
    \end{tabular}
    }
\end{table}

\subsection{Ablation Study}

We conduct ablation studies to validate the effectiveness of our multi-condition CFG strategy. 
As shown in Table~\ref{tab:3}, increasing the factor \(s_t\) improves WER, but at the cost of lip-speech synchronization. The factor \(s_f\) and \(s_l\) directly affects lip synchronization. Specifically, facial features (\(s_f\)) serve as a weak synchronization cue, causing minimal disruption to speech quality. In contrast, lip features (\(s_l\)) enforce stronger synchronization, improving lip-speech alignment but slightly affecting WER. These results reveal a trade-off between synchronization and speech quality, suggesting that optimizing one objective may come at the cost of the other.

\section{Conclusion}

In this work, we propose a novel automatic video dubbing method that improves both lip synchronization and speech quality compared to baseline approaches. We also explore its application to bilingual scenarios, demonstrating its effectiveness in both English and Chinese. Future work will focus on further enhancing dubbing performance, particularly by expanding to more languages and improving cross-lingual adaptation.

\section{Acknowledgements}

This work was supported in part by the National Natural Science Foundation of China under Grants 62276220 and 62371407 and the Innovation of Policing Science and Technology, Fujian province (Grant number: 2024Y0068).


\bibliographystyle{IEEEtran}

\bibliography{mybib}


\end{document}